\documentclass[conference]{IEEEtran}

\IEEEoverridecommandlockouts

\usepackage{cite}
\usepackage{amsmath,amssymb,amsfonts}
\usepackage{tikz}
\usetikzlibrary{patterns}
\usetikzlibrary{decorations.pathreplacing}
\usetikzlibrary{shapes.geometric}
\usepackage{pgfplots}
\usetikzlibrary{calc}
\usetikzlibrary{arrows.meta}
\usepackage{amsmath}
\usepackage{amsfonts}
\usepackage{amssymb}
\usepackage{mathtools}
\usepackage{algorithmic}
\usepackage{graphicx}
\usepackage{textcomp}
\usepackage{blkarray}
\usepackage{array}
\usepackage{multirow}
\usepackage{arydshln}

\usepackage{cite}
\usepackage{amsmath,amssymb,amsfonts}
\usepackage{tikz}
\usetikzlibrary{patterns}
\usetikzlibrary{decorations.pathreplacing}
\usetikzlibrary{matrix,decorations.pathreplacing, calc, positioning}
\usetikzlibrary{shapes.geometric}
\usepackage{pgfplots}
\usetikzlibrary{calc}
\usetikzlibrary{arrows.meta}
\usepackage{amsmath}
\usepackage{amsfonts}
\usepackage{arydshln}
\usepackage{amssymb}
\usepackage{mathtools}
\usepackage{algorithmic}
\usepackage{graphicx}
\usepackage{textcomp}
\usepackage{blkarray}
\usepackage{array}
\usepackage{multirow}
\usetikzlibrary{patterns}

\newcommand\numberthis{\addtocounter{equation}{1}\tag{\theequation}}
\usepackage{xcolor}
\def\BibTeX{{\rm B\kern-.05em{\sc i\kern-.025em b}\kern-.08em
    T\kern-.1667em\lower.7ex\hbox{E}\kern-.125emX}}
    
\makeatletter
\newcommand*{\rom}[1]{\expandafter\@slowromancap\romannumeral #1@}
\makeatother

\begin{document}

\title{\huge Age of Actuation in a Wireless Power Transfer System}

\author{
	\IEEEauthorblockN{Ali Nikkhah\IEEEauthorrefmark{1}, 
	Anthony Ephremides\IEEEauthorrefmark{2}, and Nikolaos Pappas\IEEEauthorrefmark{1}\\}
    \IEEEauthorblockA{\IEEEauthorrefmark{1}Department of Computer and Information Science, Link\"{o}ping University, Link\"{o}ping, Sweden}
    \IEEEauthorblockA{\IEEEauthorrefmark{2}Electrical and Computer Engineering, University of Maryland, College Park, MD, USA}
   \{ali.nikkhah, nikolaos.pappas\}@liu.se, etony@umd.edu
   \thanks{This work has been supported in part by the Swedish Research Council (VR), ELLIIT, Zenith, and the European Union (ETHER, 101096526).}
}

\maketitle
\begin{abstract} 
In this paper, we study a model relevant to semantics-aware goal-oriented communications. More specifically, observations from an external process are transmitted through status updates to a battery-powered receiver. From these updates, the receiver is informed about the status of the process and if there is sufficient energy, uses them to perform an actuation to achieve a goal. We consider a wireless power transfer model where the destination receives energy from a dedicated power transmitter and occasionally from the data transmitter. We provide the analysis for the Age of Information (AoI). Furthermore, we propose a new metric, namely the \textit{Age of Actuation (AoA) which is relevant when the receiver utilizes the status updates to perform actions in a timely manner}. We analytically characterize the AoA and we show that is a more general metric than AoI. We provide the optimization problems for both metrics, and we numerically evaluate our analytical results.
\end{abstract}

\vspace{-5pt}

\section{Introduction}

In the current communication systems, the importance and effectiveness of information have been disregarded. Semantics-aware goal-oriented communication \cite{kountouris2021semantics, gunduz2022beyond, Timing6G} goes towards the direction of incorporating the semantics of information in the whole information chain from its generation to its utilization. Age of Information (AoI) \cite{TutorialYates, Pappas2023age} is a metric that captures the timeliness of information which is a semantic attribute in status-updating systems. However, AoI cannot capture the utilization of information on the timeliness of a data-based action. In this work, we define a new metric namely the \textit{Age of Actuation (AoA)} that captures the elapsed time since the last performed actuation at a remote destination based on data received by a source. In order to perform the actuation, the destination needs energy which is provided wirelessly by a power-transmitting device.

Radio frequency wireless power transfer (WPT) \cite{varshney2008transporting} is important in networks with energy harvesting (EH) devices that cannot be connected to the power grid and they are hard to reach to have replaceable batteries. Several studies consider the timeliness of information in EH systems. In the vast majority of these studies, energy consumption is related to the transmission of information instead of utilizing the information for performing an action. There are two types of EH systems, first, the EH from the environment is modeled as an external stochastic process. More specifically, in works where time is assumed to be continuous, the process of EH is modeled by a Poisson process \cite{YatesISIT2015, feng2018optimal,zheng2019closed,bacingolu2019optimal, ArafaTWC2019, elmagid2022age}. In discrete-time models, EH is modeled by a Bernoulli process\cite{pappas2020average,hatami2021aoi,chen2021optimization, CeranAoI2019, xie2022age}.

The second type of EH is when the energy is harvested by WPT. The works \cite{krikidis2019average, khorsandmanesh2021average, abd2020aoi, leng2019ageof, Abd-ElmagidTCOM2020} consider WPT in several setups with the utilization of different tools for optimization such as the Markov Decision Processes framework and the Reinforcement Learning. The aforementioned papers though consider the utilization of harvested energy to transmit information, they do not consider how to perform an action based on the received information and the required energy for the action.

In this work, a battery-powered receiver utilizes a received status update to be informed about the status of a remote process and to perform an actuation. The reception of an update does not consume energy, but the actuation does. There is a dedicated node that transmits wireless power and a node that monitors an external process and transmits the status update to the receiver according to the generate-at-will policy. Both nodes share the same wireless channel hence the power transmission interferes with the data transmission. Our model belongs to the Simultaneous Wireless Information and Power Transfer (SWIPT) \cite{zhang2013mimo}.

The contributions of this paper are as follows:

\begin{itemize}
\item We consider a model relevant for semantics-aware goal-oriented communications where a device generates status updates that inform the battery-powered remote destination about an external process. The destination can utilize these updates to perform an action to achieve a goal. We consider a SWIPT model where the destination receives energy from a dedicated transmitter and occasionally from the data transmitter. 

\item We provide the analysis for the AoI at the receiver. Furthermore, we characterize analytically the evolution of the battery and we utilize these results for the characterization of the AoA that is defined in this work. \textit{AoA as we show, is a more general metric than AoI and becomes relevant when the receiver uses the status updates to perform actions in a timely manner which is crucial for goal-oriented communications}.
\item We consider two optimization problems namely the minimization of the average AoI and the minimization of the average AoA.
\item We provide numerical evaluation of the analytical results.
\end{itemize}

\section{System Model} \label{System Model}
\begin{figure}[b]
\centering
\resizebox{0.48\textwidth}{!}{
\begin{tikzpicture}

\draw [thick] (0,0) rectangle (2,1) node[scale=0.8] [xshift=-1.23cm,yshift=0.3cm] {Transmitter 1}; 
\draw [thick] (0,0.7) to [bend right=-35] (0.2,0.7) to [bend right=-15]  (0.4,0.1) to [bend right=5] (0.5,0.65) -- (0.7,0.9) -- (0.8,0.8) -- (0.85,0.92) to [bend right=10] (1,0.2)-- (1.05,0.27) -- (1.11,0.08) --(1.14,0.52) -- (1.2,0.47) -- (1.3,0.8) -- (1.5,0.2) -- (1.7,0.3) -- (1.78,0.15) -- (1.88,0.9) -- (2,0.2);

\draw [thick] (2,0.5) -- (2.5,0.5) -- (3,1);
\draw[thick] (3,0.5) -- (4,0.5) -- (3.7,1) -- (4.3,1) -- (4,0.5);
\draw [-Stealth,thick] (2.5,1) to [bend right=-45] (3,0.5) node [xshift=-0.4cm, yshift=0.7cm] {$q_1$};
\draw [-Stealth, thick, dashed] (4.35,0.85) -- (5.6,0.3);
\draw [-Stealth, thick] (4.35,0.65) -- (5.6,0.1);

\draw[thick] (4,-2.4) rectangle (6.8,-1.4) node[xshift=-1.8cm,yshift=-0.3cm, scale=0.8] {Data Flow} node[xshift=-1.8cm,yshift=-0.7cm, scale=0.8] {Energy Flow} ;
\draw[-Stealth,thick, dashed] (6,-1.7) -- (6.7,-1.7);
\draw[-Stealth,thick] (6,-2.1) -- (6.7,-2.1);

\node[scale=0.8] [xshift=1.25cm,yshift=-2.5cm] {Transmitter 2};

\draw[fill=black, thin, scale=1, xshift=1cm , yshift=-1cm] (0,0) -- (45:0.4cm) -- (15:0.35cm) -- (45:1cm) -- (45:0.6cm) -- (60:0.65);

\begin{scope}[yscale=1,xscale=-1]
\draw[fill=black, thin, scale=1, xshift=-1cm , yshift=-1cm] (0,0) -- (45:0.4cm) -- (15:0.35cm) -- (45:1cm) -- (45:0.6cm) -- (60:0.65);
\end{scope}

\begin{scope}[yscale=-1,xscale=1]
\draw[fill=black, thin, scale=1, xshift=1cm , yshift=1cm] (0,0) -- (45:0.4cm) -- (15:0.35cm) -- (45:1cm) -- (45:0.6cm) -- (60:0.65);
\end{scope}

\begin{scope}[yscale=-1,xscale=-1]
\draw[fill=black, thin, scale=1, xshift=-1cm , yshift=1cm] (0,0) -- (45:0.4cm) -- (15:0.35cm) -- (45:1cm) -- (45:0.6cm) -- (60:0.65);
\end{scope}

\draw[thick] (1,-1) -- (2.5,-1) -- (3,-0.5);
\draw[thick] (3,-1) -- (4,-1) -- (3.7,-0.5) -- (4.3,-0.5) -- (4,-1);
\draw [-Stealth, thick] (2.5,-0.5) to [bend right=-45] (3,-1) node [xshift=-0.4cm, yshift=0.7cm] {$q_2$};
\draw [-Stealth, thick] (4.35,-0.75) -- (5.6,0);

\draw[thick] (6,0) -- (5.7,0.5) -- (6.3,0.5) -- (6,0) -- (6.5,0);

\draw[thick] (6.5,0) -- (7,1) -- (7.5,1);
\draw[thick] (7.75,1) circle (0.25cm);
\draw[thick] (7.75,1) -- (7.93,1.18);
\draw[thick] (7.75,1) -- (7.93,0.82);
\draw[thick] (7.75,1) -- (7.57,1.18);
\draw[thick] (7.75,1) -- (7.57,0.82);

\draw[-Stealth] (7.75,1.75) -- (7.75,1.3) node [xshift=0cm, yshift=0.75cm] {$\sqrt{1-\rho^2}$};
\draw[-Stealth, thick] (8,1) -- (8.5,1);

\begin{scope}[yscale=-1,xscale=1]
\draw[thick] (6.5,0) -- (7,1) -- (7.5,1);
\draw[thick] (7.75,1) circle (0.25cm);
\draw[thick] (7.75,1) -- (7.93,1.18);
\draw[thick] (7.75,1) -- (7.93,0.82);
\draw[thick] (7.75,1) -- (7.57,1.18);
\draw[thick] (7.75,1) -- (7.57,0.82);

\draw[-Stealth, thick] (7.75,1.75) -- (7.75,1.3) node [xshift=0cm, yshift=-0.75cm] {$\rho$};
\draw[-Stealth, thick] (8,1) -- (8.5,1);
\end{scope}

\draw [thick] (8.5,1.5) rectangle (9.5,-1.5);
\draw [thick] (8.5,0) -- (9.5,0) node [xshift=-0.5cm, yshift=0.75cm] {ID} node [xshift=-0.5cm, yshift=-0.75cm] {EH} node[scale=0.8] [xshift=-2.2cm, yshift=0cm] {Receiver};

\draw[-Stealth, thick] (9.5,1) -- (11,1);
\draw[fill,black] (11.2,1) circle (0.2cm);
\draw[thick] (11.25,1.05) -- (11.25-0.7,1.05+0.7);
\draw[thick] (11.15,0.95) -- (11.15-0.7,0.95+0.7);
\draw[fill,black] (10.50,1.70) circle (0.11cm);
\node [ xshift=12cm, yshift=1.70cm,scale=0.95] {Actuation};
\draw[thick] (10.55,1.65) -- (10.55+0.7,1.65+0.7);
\draw[thick] (10.45,1.75) -- (10.45+0.7,1.75+0.7);

\draw[fill=black] (10.55+0.8,1.65+0.6) -- (10.45+0.6,1.75+0.8) -- (10.55+0.8,1.75+0.8)-- (10.55+0.8,1.65+0.6);

\draw[-Stealth, thick] (9.5,-1) -- (10,-1);
\draw[thick] (10,-0.5) rectangle (13,-1.5);
\draw[fill=gray] (12,-0.5) rectangle (13,-1.5);
\draw (12.5,-1.5) -- (12.5,-0.5);
\draw (12,-1.5) -- (12,-0.5);
\draw (11.5,-1.5) -- (11.5,-0.5);
\draw (10.5,-1.5) -- (10.5,-0.5);

\draw[fill,black] (10.75,-1) circle (0.05cm);
\draw[fill,black] (11,-1) circle (0.05cm);
\draw[fill,black] (11.25,-1) circle (0.05cm);

\draw[-Stealth, thick] (12.75,-0.5) -- (11.34,0.86);

\end{tikzpicture}
}
\caption{The considered system model.} 

\label{SystemModelFig}
\end{figure}
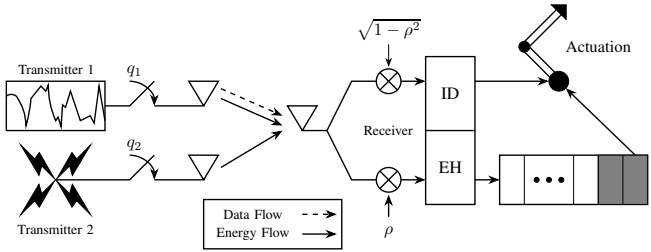

\vspace{-4pt}
We consider a system consisting of two transmitting devices as depicted in Fig. \ref{SystemModelFig}; the first one is transmitting status updates as data packets to a receiver, and the second is dedicated to power transmission. The devices share the same wireless channel under random access. The receiver performs an actuation upon reception of a data packet, given that there is available stored energy in its battery in terms of energy packets. The transmitters provide the receiver with the necessary data and energy for performing an actuation. Thus, there is SWIPT, which consists of energy harvesting (EH) and information decoding (ID) units, at the receiver side. The parameter $\rho^2$, as illustrated in Fig. \ref{SystemModelFig}, denotes the power splitting ratio. More specifically, $\rho^2$ and $1-\rho^2$ are the fractions of split and directed power to the EH and the ID circuits, respectively. Time is discrete and divided into timeslots, $t$, of one time unit each. The first transmitter observes a random process and samples and transmits a status update with probability (w.p.) $q_1$ in a timeslot. This status update will inform the receiver about the status of the remote source, and then it will be utilized to perform an action if there is sufficient energy.

The receiver can harvest energy from both transmitters as explained below. The second transmitter transmits wireless power w.p. $q_2$ in a time slot. The first transmitter, while transmitting data, can also assist with the energy only when the second transmitter transmits. In the case of a successful energy transmission, an energy packet is stored in the battery. More details are given later. We consider the cases of finite-sized and infinite-sized batteries.

At the end of a time slot, the receiver can perform an actuation by utilizing one energy packet from its battery and using the received data packet in the same time slot. In case there is no data reception, or not sufficient energy, the receiver cannot perform the action. Thus, an actuation is performed in a time slot according to the following two cases: i) there is a non-empty battery, and there is a successful data reception, ii) there is an empty battery, but there are successful receptions of both energy and data packets during that time slot.

\subsection{Physical Layer Model} \label{Physical Layer Model}

We denote $\cal{D}$ the event of a successful transmission of a data packet in a time slot which occurs w.p.

\begin{equation} \label{P_u1}
P_{\cal{D}}=q_1q_2P_{d12}+q_1\bar{q}_2P_{d1},
\end{equation}
where $P_{d1}$ is the success probability for the data packets when only the first transmitter attempts to transmit; $P_{d12}$ is the success probability when both transmitters transmit simultaneously. The expressions are given in Section \ref{The Success Probabilities of Transmissions in Different Scenarios}. We assume that if a packet is not successfully transmitted in a slot then it is dropped by the transmitter. In this study we do not require the existence of a feedback channel from the receiver to the transmitter.

We denote $\cal{E}$ the event of a successful transmission of an energy packet in a time slot, it occurs w.p.
\vspace{-6pt}
\begin{equation} \label{P_u2}
P_{\cal{E}}=q_1q_2P_{e12}+\bar{q}_1q_2P_{e2},
\vspace{-3pt}
\end{equation}
where $P_{e12}$, $P_{e2}$ are the probabilities of a successful transmission of an energy packet when both of the transmitters transmit and when only the second transmitter transmits, respectively. 

The expressions are given in Section \ref{The Success Probabilities of Transmissions in Different Scenarios}.

The probabilities of the four possible joint outcomes are
\vspace{-8pt}
\begin{align}
 \label{P_u1_u2}
&P_{\cal{D,E}}=q_1 q_2 P_{d12} P_{e12},\\ 
 \label{P_u1_u2b}
&P_{\cal{D,\bar{E}}}= q_1 q_2 P_{d12} \bar{P}_{e21} +q_1 \bar{q}_2 P_{d1},\\
\label{P_u1b_u2}
&P_{\cal{\bar{D},E}}=q_1 q_2 \bar{P}_{d12} P_{e12} + \bar{q}_1 q_2 P_{e2},\\
\label{P_u1b_u2b}
&P_{\cal{\bar{D},\bar{E}}}=\bar{q}_1\bar{q}_2+q_1 \bar{q}_2 \bar{P}_{d1} + \bar{q}_1 q_2 \bar{P}_{e2} + q_1 q_2 \bar{P}_{d12} \bar{P}_{e12}.
\end{align}

Note that $P_{\cal{D}}=P_{\cal{D,E}}+P_{\cal{D,\bar{E}}}$ and $P_{\cal{E}}=P_{\cal{D,E}}+P_{\cal{\bar{D},E}}$.

\subsection{Success probabilities of data and energy transmissions} \label{The Success Probabilities of Transmissions in Different Scenarios}

Here, we provide more details on how to obtain the success probabilities used in the previous expressions. We assume that the transmitting power by the $j$-th transmitter is denoted by $P_{tx,j}$, $j \in \{1,2\}$. Thus $P_{rx,j}= g_j A_j$ is the received power by the $j$-th transmitter, where $g_j=P_{tx,j} d_j^{-\alpha_j}$, $d_j$ is the distance between transmitter $j$ and the receiver, and $\alpha_j$ is the path loss exponent from the $j$-th transmitter to the receiver. $A_j$ denotes the small-scale fading for the link of the $j$-th transmitter and in this work we assumed that is Rayleigh fading.

We consider a successful transmission of data when the signal to interference and noise ratio (SINR) of the data transmission is above a threshold, i.e., $SINR \geq \gamma_d$. We also consider a successful transmission of an energy packet whenever the harvested energy at the receiver is above a threshold, i.e., $E \geq \gamma_e$.
The probabilities of successful transmissions of data and energy packets can be characterized as follows.
\subsubsection{Both Transmitters Transmit ($0 \leq \rho \leq 1$)} \label{Both Transmitters Transmit}
A data transmission is successful when $SINR_{d12}=\frac{P_{rx,1}}{P_{rx,2}+\frac{P_N}{1-\rho^2}} \geq \gamma_d$ and it occurs w.p.
\vspace{-5pt}
\begin{equation} \label{T_d12}
\displaystyle{P_{d12}=\text{exp}\left(-\frac{\gamma_d  P_N}{(1-\rho^2) \upsilon_1  g_1}\right) \left(1+\gamma_d \frac{\upsilon_2 g_2}{\upsilon_1 g_1}\right)^{-1}}.
\end{equation} 
This can be obtained by \cite[Theorem 1]{nguyen2008optimization}, dividing the noise power by $1-\rho^2$.
A successful transmission of an energy packet occurs when
$E_{e12}= \rho^2(P_{rx,1} +  P_{rx,2}) \geq \gamma_e$, w.p.
\begin{equation}
\resizebox{0.5\textwidth}{!}{$
\displaystyle{P_{e12}}=
1-\frac{g_2 \upsilon_2 (\text{exp}(-\frac{\gamma_e}{\rho^2 g_2 \upsilon_2})-1)+g_1 \upsilon_1 (1-\text{exp}(-\frac{\gamma_e}{\rho^2 g_1 \upsilon_1}))}{g_1\upsilon_1 - g_2 \upsilon_2}.$ \numberthis \label{T_e12}}
\vspace{-5pt}
\end{equation}
The proof is omitted due to space limitations.

\subsubsection{Only the First Transmitter Transmits ($ \rho = 0$)} \label{Only the First Transmitter Transmits}
The probability of a successful data transmission in this case is given by $\displaystyle{P_{d1}=\text{exp}\left(-\frac{\gamma_d P_N}{\upsilon_1  g_1} \right)}$.

\subsubsection{Only the Second Transmitter Transmits ($\rho=1$)} \label{Only the Second Transmitter Transmits}
The probability of a successful energy transmission 
is given by $P_{e2}= \text{exp} (- \frac{\gamma_e}{ g_2 \upsilon_2})$, and it is obtained from (\ref{T_e12}), after setting $g_1=0$ and $\rho=1$.

\section{Analytical Results} \label{Analytical Results}
In this section, we present the analysis for the evolution of the battery, the AoI, and the proposed metric of Age of Actuation (AoA). 
\subsection{Age of Information (AoI)} \label{The Evolution of Age of Information}

The reception of data packets is independent of the energy availability. Here we provide a brief discussion on AoI metric. We consider the generate-at-will policy, thus, at time slot $t+1$, the value of the AoI denoted by $I(t+1)$ drops to $1$, if there is a successful data reception, otherwise AoI, $I(t+1)$, increases by $1$, a sample path of AoI is depicted in vertical lines by Fig. \ref{AoA_AoI_Evo}. The AoI evolution is summarized by

\begin{equation} \label{AoI_evo}
I(t+1)=\begin{cases}
1 & \text{w.p.} \  P_{\cal{D}} , \\
I(t)+1 & \text{w.p.} \  1-P_{\cal{D}} . \\
\end{cases}
\end{equation}
AoI can be modelled as a Discrete Time Markov Chain (DTMC)\cite{fountoulakis2022information}, and the average AoI is given by $\bar{I}=\frac{1}{P_{\cal{D}}}$. Note that we can also obtain its stationary distribution as in \cite{fountoulakis2022information}. Details are omitted due to space limitations.

\subsection{The Evolution of the Battery} \label{The Evolution of the Battery}

Before proceeding with the analysis for the AoA, we need to provide some analytical results on the battery.
We consider two cases, the finite-sized and the infinite-sized batteries. We denote the state of the battery at a time slot $t$ by $B(t)$, then we have $B(t)=\{0,1,\hdots,m\}$ and $B(t)=\{0,1,\hdots,\infty\}$ for two cases of finite-sized ($m$ is the size of the battery) and infinite-sized batteries, respectively. The evolution of the battery can be described as follows.
If $B(t)=0$ then
\vspace{-5pt}
\begin{equation} \label{B_evo_first}
B(t+1)=\begin{cases}
0 & \text{w.p.} \ P_{\cal{D,E}}+ P_{\cal{D,\bar{E}}}+P_{\cal{\bar{D},\bar{E}}} , \\
1 & \text{w.p.} \  P_{\cal{\bar{D},E}} . \\
\end{cases}
\end{equation}
In the finite case if $ 1 \leq B(t) \leq m-1$, and in the infinite case if $B(t) \geq 1$, we have
\begin{equation} \label{B_evo_middle}
B(t+1)=\begin{cases}
B(t)-1 & \text{w.p.} \  P_{\cal{D,\bar{E}}} , \\
B(t) & \text{w.p.} \  P_{\cal{D,E}}+P_{\cal{\bar{D},\bar{E}}} , \\
B(t)+1 & \text{w.p.} \  P_{\cal{\bar{D},E}} . \\
\end{cases}
\end{equation}
If $B(t)=m$, in the finite-sized battery we have
\begin{equation}  \label{B_evo_last}
B(t+1)=\begin{cases}
m-1 & \text{w.p.} \  P_{\cal{D,\bar{E}}} , \\
m & \text{w.p.} \  P_{\cal{D,E}}+P_{\cal{\bar{D},E}}+P_{\cal{\bar{D},\bar{E}}} . \\
\end{cases}
\end{equation}

The battery size evolution can be modelled by a DTMC for both cases. The transition probabilities are omitted due to space limitations but they can be obtained by the previous expressions. Their steady state distributions are represented by $\pi$, and $\pi_k$ is the probability that the battery has stored $k$ energy packets. The probability that the battery is empty $\mathrm{Pr}(B=0)$ is equal to $\pi_0$ and we denote $\bar{\pi}_0$ the probability of a non-empty battery. To simplify the notation we denote the event of a non-empty battery by ${\mathcal{B}}$. The expressions for the aforementioned probabilities are provided below.

The infinite-sized and finite-sized batteries, can be modelled by a Geo/Geo/1 queue and a Geo/Geo/1/B queue, respectively \cite[Sections 3.4.2 and 3.4.3]{srikant2013communication}. In \cite{srikant2013communication}, service utilization is defined as the probability of one arrival and no departure divided by the probability of one departure and no arrival. One can see that the former probability is $P_{\cal{\bar{D},E}}$ and the latter probability is $P_{\cal{D,\bar{E}}}$. Hence, for infinite sizes, the probability of an empty battery is $\mathrm{Pr}(B=0)=1-\frac{P_{\bar{{\cal{D}}},{\cal{E}}}}{P_{{\cal{D}},\bar{{\cal{E}}}}}$ and a non-empty battery is $\mathrm{Pr}(B\neq 0)=\frac{P_{\bar{{\cal{D}}},{\cal{E}}}}{P_{{\cal{D}},\bar{{\cal{E}}}}}$, respectively.
Note that, for the infinite-sized battery, these are valid relations when the stability condition, $P_{\bar{{\cal{D}}},{\cal{E}}} < P_{{\cal{D}},\bar{{\cal{E}}}}$, holds. When $P_{\bar{{\cal{D}}},{\cal{E}}} \geq P_{{\cal{D}},\bar{{\cal{E}}}}$, then the battery never empties which is equivalent to an always connected to the power grid system. Thus we have
\vspace{-5pt}
\begin{equation} \label{empty_p_inf}
\pi_0^i=
\begin{cases}
0 & \frac{P_{\bar{{\cal{D}}},{\cal{E}}}}{P_{{\cal{D}},\bar{{\cal{E}}}}} \geq 1, \\
1- \frac{P_{\bar{{\cal{D}}},{\cal{E}}}}{P_{{\cal{D}},\bar{{\cal{E}}}}}  & \frac{P_{\bar{{\cal{D}}},{\cal{E}}}}{P_{{\cal{D}},\bar{{\cal{E}}}}} < 1, \\
\end{cases}
\quad
\bar{\pi }_0^i=
\begin{cases}
1 & \frac{P_{\bar{{\cal{D}}},{\cal{E}}}}{P_{{\cal{D}},\bar{{\cal{E}}}}} \geq 1, \\
\frac{P_{\bar{{\cal{D}}},{\cal{E}}}}{P_{{\cal{D}},\bar{{\cal{E}}}}} & \frac{P_{\bar{{\cal{D}}},{\cal{E}}}}{P_{{\cal{D}},\bar{{\cal{E}}}}} < 1. \\
\end{cases}
\end{equation}

For the finite-sized battery, the probabilities of empty and non-empty batteries are
\vspace{-5pt}
\begin{equation} \label{empty_p_fin}
\pi_0^f=\frac{1-\frac{P_{\bar{{\cal{D}}},{\cal{E}}}}{P_{{\cal{D}},\bar{{\cal{E}}}}}}{1-\left(\frac{P_{\bar{{\cal{D}}},{\cal{E}}}}{P_{{\cal{D}},\bar{{\cal{E}}}}}\right)^{m+1}} , 
\quad
\bar{\pi}_0^f=\frac{\frac{P_{\bar{{\cal{D}}},{\cal{E}}}}{P_{{\cal{D}},\bar{{\cal{E}}}}}-\left(\frac{P_{\bar{{\cal{D}}},{\cal{E}}}}{P_{{\cal{D}},\bar{{\cal{E}}}}}\right)^{m+1}}{1-\left(\frac{P_{\bar{{\cal{D}}},{\cal{E}}}}{P_{{\cal{D}},\bar{{\cal{E}}}}}\right)^{m+1}} .  
\end{equation}
\vspace{-2pt} 
We can see that (\ref{empty_p_fin}) transform to (\ref{empty_p_inf}) as $m \rightarrow \infty$, when $\frac{P_{\bar{{\cal{D}}},{\cal{E}}}}{P_{{\cal{D}},\bar{{\cal{E}}}}} < 1$. We use the superscripts $i$ and $f$, to distinguish between \textit{infinite}-sized and \textit{finite}-sized batteries, respectively. 
\subsection{Age of Actuation (AoA)} \label{The Evolution of Actuation Freshness}

\begin{figure}
\centering 
\scalebox{.85}{ \boldmath{
\begin{tikzpicture}

\fill[pattern=horizontal lines] (0,0) rectangle (1,1);
\fill[pattern=horizontal lines] (1,0) rectangle (2,2);
\fill[pattern=horizontal lines] (2,0) rectangle (3,3);
\fill[pattern=horizontal lines] (3,0) rectangle (4,4);
\fill[pattern=horizontal lines] (4,0) rectangle (5,5);
\fill[pattern=horizontal lines] (5,0) rectangle (6,1);
\fill[pattern=horizontal lines] (6,0) rectangle (7,2);

\fill[pattern=vertical lines] (0,0) rectangle (1,1);
\fill[pattern=vertical lines] (1,0) rectangle (2,2);
\fill[pattern=vertical lines] (2,0) rectangle (3,1);
\fill[pattern=vertical lines] (3,0) rectangle (4,2);
\fill[pattern=vertical lines] (4,0) rectangle (5,3);
\fill[pattern=vertical lines] (5,0) rectangle (6,1);
\fill[pattern=vertical lines] (6,0) rectangle (7,2);

\draw[-Stealth, very thick ] (0,0) -- (9,0)  node [xshift=0cm, yshift=-0.5cm] {$t$} node [xshift=-9cm, yshift=-0.5cm] {$0$} node [xshift=-7cm, yshift=-0.7cm] {$({\mathcal{D}},\bar{{\mathcal{E}}} \cap \bar{{\mathcal{B}}} )$} node [xshift=-4cm, yshift=-0.7cm] {$({\mathcal{D}},{\mathcal{E}} \cup {\mathcal{B}} )$};

\draw [decorate,decoration={brace,amplitude=5pt,raise=0.5ex}, thick] (1,-0.5) --  (3,-0.5);
\draw [decorate,decoration={brace,amplitude=5pt,raise=0.5ex}, thick] (4,-0.5) --  (6,-0.5);

\draw[-Stealth, very thick ] (0,0) -- (0,6) node [xshift=-0.7cm, yshift=0cm] {$A(t)$} node [xshift=-0.7cm, yshift=-0.6cm] {$I(t)$} node [xshift=-0.4cm, yshift=-5cm] {$1$} node [xshift=-0.4cm, yshift=-6cm] {$0$};

\draw (1,-0.1) -- (1,0.1);
\draw (2,-0.1) -- (2,0.1);
\draw (3,-0.1) -- (3,0.1);
\draw (4,-0.1) -- (4,0.1);
\draw (5,-0.1) -- (5,0.1);
\draw (6,-0.1) -- (6,0.1);
\draw (7,-0.1) -- (7,0.1);
\draw (8,-0.1) -- (8,0.1);

\draw (-0.1,1) -- (0.1,1);
\draw (-0.1,2) -- (0.1,2);
\draw (-0.1,3) -- (0.1,3);
\draw (-0.1,4) -- (0.1,4);
\draw (-0.1,5) -- (0.1,5);
\draw (6,-0.1) -- (6,0.1);

\draw[very thick] (0,1) -- (1,1);

\draw[very thick] (1,1) -- (1,2);
\draw[very thick] (1,2) -- (2,2);

\draw[very thick] (2,2) -- (2,3);
\draw[very thick] (2,3) -- (3,3);

\draw[very thick] (3,3) -- (3,4);
\draw[very thick] (3,4) -- (4,4);

\draw[very thick] (4,4) -- (4,5);
\draw[very thick] (4,5) -- (5,5);

\draw[ very thick] (5,5) -- (5,1);
\draw[very thick] (5,1) -- (6,1);

\draw[ very thick] (6,1) -- (6,2);
\draw[ very thick] (6,2) -- (7,2);

\draw[dashed, very thick] (2,2) -- (2,1);
\draw[dashed, very thick] (2,1) -- (3,1);

\draw[dashed, very thick] (3,1) -- (3,2);
\draw[dashed, very thick] (3,2) -- (4,2);

\draw[dashed, very  thick] (4,2) -- (4,3);
\draw[dashed, very thick] (4,3) -- (5,3);

\node[xshift=6.5cm,yshift=5cm] {AoA};
\node[xshift=6.5cm,yshift=4.5cm] {AoI};

\fill[pattern=horizontal lines] (7,4.8) rectangle (8,5.2);
\fill[pattern=vertical lines] (7,4.3) rectangle (8,4.7);

\draw (6,4.2) rectangle (8.2,5.3);

\end{tikzpicture}}} 

\caption{The evolution of AoA and AoI metrics.}
\label{AoA_AoI_Evo}
\end{figure}
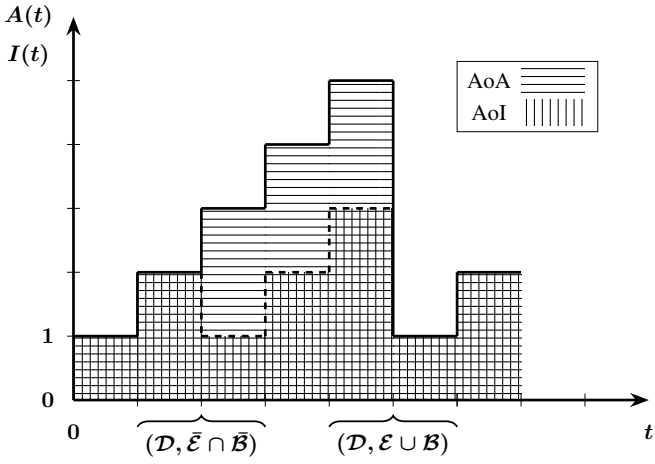

In this section we first define and then analyze a new metric that becomes relevant when status updates are used to perform actions in a timely manner. AoA, $A(t)=t-a(t)$ is the time elapsed since the last actuation was performed, where $a(t)$ denotes the time of the last performed actuation and we assume that $A(0)=1$. In Fig. \ref{AoA_AoI_Evo} we depict a sample path of AoA. The area of vertical lines represent the AoI, the area of horizontal lines depict the AoA, thus, the grid area is the overlap. We observe that at $t=2$ and $t=5$ there is a successful transmission of a data packet and the AoI is reset to $1$. However, at $t=2$, both the energy transmission is unsuccessful and the battery is empty, thus, the AoA continues to grow. At $t=5$, either the energy transmission is successful or the battery is non-empty, hence, AoA is reset to $1$.

In a time slot $t+1$, $A(t+1)=1$ when in time $t$ happened one of the following two cases. First, when there is a successful transmission of data, i.e., ${\cal{D}}$, and also the battery is non-empty, i.e., ${\mathcal{B}}$, thus it can provide the energy for an actuation, $({\mathcal{D}},{\mathcal{B}})$, it occurs w.p. $P_{\cal{D}} \bar{\pi}_0^b$. Second, if the battery is empty, i.e., $\bar{{\mathcal{B}}}$, then we need $({\cal{D}},{\cal{E}})$ meaning that we have simultaneously successful energy and data transmissions. It occurs w.p. $P_{{\cal{D}},{\cal{E}}} \pi_0^b$. Otherwise, $A(t+1)=A(t)+1$. The evolution of the AoA can be summarized by
\vspace{-4pt}
\begin{equation} \label{AF_evo}
A(t+1)=\begin{cases}
1 & \text{w.p.} \  P_{\cal{D}} \bar{\pi}_0^b + P_{\cal{D},\cal{E}} \pi_0^b , \\
A(t)+1 & \text{w.p.} \  1-(P_{\cal{D}} \bar{\pi}_0^b + P_{\cal{D},\cal{E}} \pi_0^b) . \\
\end{cases}
\vspace{-4pt}
\end{equation}
The evolution is similar to (\ref{AoI_evo}) except we have $P_{\cal{D}} \bar{\pi}_0^b + P_{\cal{D},\cal{E}} \pi_0^b$ instead of $P_{\cal{D}}$. Hence, we have

\begin{equation} \label{AF_mean_2}
\bar{A}=\frac{1}{P_{\cal{D}} \bar{\pi}_0^b + P_{\cal{D},\cal{E}} \pi_0^b}.
\vspace{-2pt}
\end{equation}
For the infinite-sized battery, by replacing $\pi_0^b$ and $\bar{\pi}_0^b$ from (\ref{empty_p_inf}), we obtain
\vspace{-8pt}
\begin{equation} \label{AF_b_inf}
\bar{A}=
\begin{cases}
\bar{A}_1=\frac{1}{P_{\cal{D}}} & \frac{P_{\bar{{\cal{D}}},{\cal{E}}}}{P_{{\cal{D}},\bar{{\cal{E}}}}} \geq 1,\\
\bar{A}_2=\frac{1}{ P_{\cal{D}} \frac{P_{\bar{{\cal{D}}},{\cal{E}}}}{P_{{\cal{D}},\bar{{\cal{E}}}}} + P_{{\cal{D}},{\cal{E}}} \left(1-\frac{P_{\bar{{\cal{D}}},{\cal{E}}}}{P_{{\cal{D}},\bar{{\cal{E}}}}}\right)  }
 & \frac{P_{\bar{{\cal{D}}},{\cal{E}}}}{P_{{\cal{D}},\bar{{\cal{E}}}}} < 1.\\
\end{cases}
\vspace{-6pt}
\end{equation}
For the finite-sized battery, similarly we obtain
\vspace{-4pt}
\begin{equation} \label{AF_b_fin}
\bar{A}=\frac{1-\left(\frac{P_{\bar{{\cal{D}}},{\cal{E}}}}{P_{{\cal{D}},\bar{{\cal{E}}}}}\right)^{m+1}}{P_{\cal{D}} \left[\frac{P_{\bar{{\cal{D}}},{\cal{E}}}}{P_{{\cal{D}},\bar{{\cal{E}}}}}-\left(\frac{P_{\bar{{\cal{D}}},{\cal{E}}}}{P_{{\cal{D}},\bar{{\cal{E}}}}}\right)^{m+1}\right]+ P_{{\cal{D}},{\cal{E}}}\left(1-\frac{P_{\bar{{\cal{D}}},{\cal{E}}}}{P_{{\cal{D}},\bar{{\cal{E}}}}}\right) } . 
\end{equation}
Note that (\ref{AF_b_fin}) transforms to (\ref{AF_b_inf}) as $m \rightarrow \infty$, when $\frac{P_{\bar{{\cal{D}}},{\cal{E}}}}{P_{{\cal{D}},\bar{{\cal{E}}}}} < 1$.

We can also obtain the stationary distribution of AoA however, due to space limitations it is omitted.

\textit{Remark 1:} When the system is not limited by energy, as we see in (\ref{AF_b_inf}), the average AoA reduces to the average AoI. This is due to the fact that there will always be available energy to perform an action, thus AoA depends only on the data arrival, as AoI. This metric is a more general case of AoI which captures a semantic attribute when an actuation is involved which is relevant in goal-oriented semantics-aware communications. 

\vspace{-10pt}

\section{Optimization} \label{Optimization}
\vspace{-5pt}
In this section we consider two optimization problems. More specifically we consider the minimization of the average AoI and the minimization of the average AoA by optimizing the transmission variables $q_1$ and $q_2$.

\vspace{-5pt}
\subsection{Optimization of the average AoI} \label{Age of Information_opt}
The AoI is not affected by battery status, thus the analysis is the same for finite-sized and infinite-sized batteries. The gradient of the objective function, $\bar{I}(q_1,q_2)=\frac{1} {{\mathcal{P}}_{D}}=\bar{A}_1$, with respect to $q_1$ and $q_2$, is
\vspace{-6pt}
\begin{align*} 
\nabla\bar{I}(q_1,q_2)=\biggr[&-\frac{P_{d1} (1 - q_2) + 
   P_{d12} q_2}{(P_{d1} q_1 (1 - q_2) + P_{d12} q_1 q_2)^2},\\
   &\frac{ q_1(P_{d1}-P_{d12}) }{(P_{d1} q_1 (1 - q_2) + P_{d12} q_1 q_2)^2}\biggr]. \numberthis \label{Grad_Inf_AoI}
\end{align*}

Given that $0 \leq q_1, q_2 \leq 1$, we observe that the first element is always negative. Also, since $P_{d1} \geq P_{d12}$, the second element is always positive. Thus, the minimum average AoI is achieved by $[q_1^{*},q_2^{*}]=[1,0]$.

and the optimum value is
\begin{equation} \label{Opt_value_AoI}
\bar{I}^{*}= \frac{1}{P_{d1}}.
\end{equation}

\textit{Remark 2:} Since AoI does not depend on energy, to minimize the average AoI we need to silence the transmitter dedicated to the power transmission to eliminate the interference. However, in a constrained optimization problem of minimizing the average AoI under average AoA constrained, this will not be the case. 
Due to space limitations, we leave this for the journal version.

\subsection{Optimization of the average AoA} \label{Optimization of the Age of Actuation}
In this section, we consider the infinite case for the battery, the objective function is (\ref{AF_b_inf}).
For the scenario where the system is not energy limited, i.e., first case in (\ref{AF_b_inf}), since $\bar{A}_1 = \bar{I}$, the analysis in Section \ref{Age of Information_opt} holds. Here we consider the $\bar{A}_2$. 
We have that
\begin{align*}
\nabla\bar{A}_2 (q_1,q_2)=\biggr[&\frac{P_{e2} - 
  P_{e12}}{((q_{1}-1) P_{e2}  -  q_{1} P_{e12})^{2} q_{2} },\\
  &\frac{1}{((q_{1}-1)P_{e2}  -q_{1}P_{e12} ) q_{2}^{2}}\biggr]. \numberthis  \label{Grad_Inf_AoA}
\end{align*}
The sign of the first element depends on $P_{e2}-P_{e12}$ and the second element is always negative.

The intersection of the two areas of $\bar{A}_1$ and $\bar{A}_2$ requires to solve $P_{\bar{{\cal{D}}},{\cal{E}}} =P_{{\cal{D}},\bar{{\cal{E}}}}$.
Then we obtain
\begin{equation} \label{Equ_Border} 
q_1  P_{d1} - q_2 P_{e2}   =  q_1 q_2( P_{d1} - P_{d12} + P_{e12} - P_{e2}). 
\end{equation}
We solve for $q_1$ (or $q_2$) and then insert it into $\bar{A}_1$ (or $\bar{A}_2$) to obtain the curve of intersection. The critical point is
\begin{equation} \label{theta1}
\resizebox{0.49\textwidth}{!}{$
q_1=\frac{P_{e2}(P_{e2}  - P_{e12}) - \sqrt{(P_{d1} - P_{d12}) P_{e2}^2 (P_{e2} - P_{e12})}}{(P_{e12} - P_{e2}) (P_{d1} - P_{d12} + P_{e12} - P_{e2})}=\theta_1,$}
\end{equation}
and
\begin{equation} \label{theta2}
\resizebox{0.5\textwidth}{!}{$
q_2=\frac{P_{d1} (P_{e12} P_{e2} - P_{e2}^2 + \sqrt{(P_{d1} - P_{d12}) P_{e2}^2 (P_{e2} - P_{e12})})}{(P_{d1} - P_{d12} + P_{e12} -  P_{e2}) \sqrt{(P_{d1} - P_{d12}) (P_{e2} - P_{e12}) P_{e2}^2}} = \theta_2.$}
\end{equation}

Also, the intersection of the two areas intersects with border of $q_1=1$ at
\begin{equation} \label{delta2}
 q_2    =  \frac{P_{d1} }{  P_{d1} +P_{e12}- P_{d12} }=\delta_2,
\end{equation} 
and intersects with the border of $q_2=1$ at
\begin{equation} \label{delta1}
q_1    =  \frac{P_{e2} }{  P_{e2} +P_{d12}- P_{e12} }= \delta_1.
\end{equation}

Comparing (\ref{Grad_Inf_AoI}) and (\ref{Grad_Inf_AoA}), if $P_{e12}>P_{e2}$, then the first elements of the gradients of $\bar{A}_1$ and $\bar{A}_2$ have the same sign, thus, the value of $\bar{A}$ would monotonously decrease with increasing of $q_1$. It can be confirmed by noting that (\ref{theta1}) and (\ref{theta2}) would be complex when $P_{e12}>P_{e2}$. Hence, $q_1^*=1$. Then, $q_2$ would be calculated by (\ref{delta2}). If $P_{e12} \geq P_{d12}$, then $\delta_2 \geq 1$ and hence $[q_1^*,q_2^*]=[1,1]$. If $P_{e12} < P_{d12}$, then $[q_1^*,q_2^*]=[1,\delta_2]$.

If $P_{e12}<P_{e2}$, then (\ref{theta1}) and (\ref{theta2}) would have rational values. Then, if $\theta_1, \theta_2 \leq 1$, we have $[q_1^*,q_2^*]=[\theta_1,\theta_2]$. Otherwise, the minimum would be on the border of either $q_1=1$ or $q_2=1$. From (\ref{delta2}) and (\ref{delta1}), it is observable that if $P_{e12} < P_{d12}$ then $[q_1^*,q_2^*]=[\delta_1,1]$, if $P_{e12} > P_{d12}$ then $[q_1^*,q_2^*]=[1,\delta_2]$, and if $P_{e12} = P_{d12}$ then $[q_1^*,q_2^*]=[1,1]$.

Thus, the $[q_1^*,q_2^*]$ that minimizes the average AoA is 
\begin{equation} \label{Opt_point_Ave_AoA}
\resizebox{0.5\textwidth}{!}{$
\begin{cases}
[1,1] & P_{e12} > P_{e2}, P_{d12} \geq P_{e12}, \\
[1,\delta_2] & P_{e12} > P_{e2}, P_{d12} < P_{e12}, \\
[\theta_1,\theta_2] & P_{e12} < P_{e2}, \{\theta_1 <1, \theta_2<1\},\\
[\delta_1,1] & P_{e12} < P_{e2}, \{\theta_1 <1, \theta_2>1\} \cup \{\theta_1 >1, \theta_2>1\}, P_{d12} > P_{e12},\\
[1,\delta_2] & P_{e12} < P_{e2}, \{\theta_1 >1, \theta_2<1\} \cup \{\theta_1 >1, \theta_2>1\}, P_{d12} < P_{e12},\\
[1,1] & P_{e12} < P_{e12} < P_{e2}, \{\theta_1 >1, \theta_2>1\}, P_{d12} = P_{e12}.
\end{cases}$}
\end{equation}

The minimum average AoA, $\bar{A}^*$, is
\begin{equation} \label{Opt_value_Ave_AoA}
\resizebox{0.5\textwidth}{!}{$
\begin{cases}
\frac{1}{P_{e12}} & P_{e12} > P_{e2}, P_{d12} \geq P_{e12}, \\
\frac{P_{d1} - P_{d12} + P_{e12}}{P_{d1} P_{e12}} & P_{e12} > P_{e2}, P_{d12} < P_{e12}, \\
\bar{I}(\theta_1,\theta_2) & P_{e12} < P_{e2}, \{\theta_1 <1, \theta_2<1\},\\
\frac{P_{d12} - P_{e12} + P_{e2}}{P_{d12} P_{e2}} & P_{e12} < P_{e2}, \{\theta_1 <1, \theta_2>1\} \cup \{\theta_1 >1, \theta_2>1\}, P_{d12} > P_{e12},\\
\frac{P_{d1} - P_{d12} + P_{e12}}{P_{d1} P_{e12}} & P_{e12} < P_{e2}, \{\theta_1 >1, \theta_2<1\} \cup \{\theta_1 >1, \theta_2>1\}, P_{d12} < P_{e12},\\
\frac{1}{P{e12}}=\frac{1}{P_{d12}} & P_{e12} < P_{e2}, \{\theta_1 >1, \theta_2>1\}, P_{d12} = P_{e12}.
\end{cases}$}
\end{equation}

\section{Numerical Results} \label{Numerical and Simulation Results}
In this section we provide numerical evaluation of the previously presented analytical results.
We consider two different setups.
For the first setup, the parameters are: $P_{tx,1}=10 \text{mW}$, $P_{tx,2}=1 \text{W}$, $P_N=-50 \text{dBm}$, $\gamma_d=\gamma_e=-10 \text{dB}$, $d_1=1 \text{m}$, $d_2=2 \text{m}$, $\alpha_1=\alpha_2=4$, $\upsilon_1=\upsilon_2=1$, and $\rho=0.99$. For the second setup, we alter only the value for $d_2=1.5 \text{m}$. The success probabilities are presented in Table \ref{Table_Setups}.
\begin{table}[b]
\begin{center}
\caption{Success probabilities of data and energy transmission.}
\label{Table_Setups}
\scalebox{1}{
\begin{tabular}{c|c|c|c|c|}
  \cline{2-5}
&   $P_{d1}$ & $P_{d12}$ & $P_{e2}$ & $P_{e12}$   \\  \cline{1-5}
\multicolumn{1}{ |c|  }{\multirow{1}{*}{Setup 1} } & 1 & 0.62 & 0.20 & 0.23  \\ \cline{1-5}
\multicolumn{1}{ |c|  }{\multirow{1}{*}{Setup 2} } & 1 & 0.34 & 0.60 & 0.63  \\ \cline{1-5}
\end{tabular}
}
\end{center}
\end{table}

\begin{figure}[h]
\centerline{\includegraphics[scale=0.42]{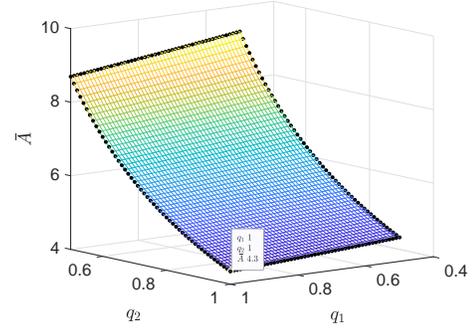}}
\vspace{-10pt}
\caption{Average AoA for the infinite-sized battery for the first setup. The minimum $\bar{A}^*=4.3$ is achieved by $q^*_1=1$ and $q^*_2=1$.}
\label{First_set_AoA}
\end{figure}

In the first/second setup receiving a data packet is more/less likely than an energy packet when both transmitters are active. \textit{Note that $P_{e12} \geq P_{e2}$ is feasible, with large values of $\rho$. This is because by power splitting, we can also utilize energy from the first transmitter.} In the figures, we show the optimal point by a data tip. Furthermore, the region of the limited energy regime is included in a black frame.

\begin{figure}[h]
\centerline{\includegraphics[scale=0.42]{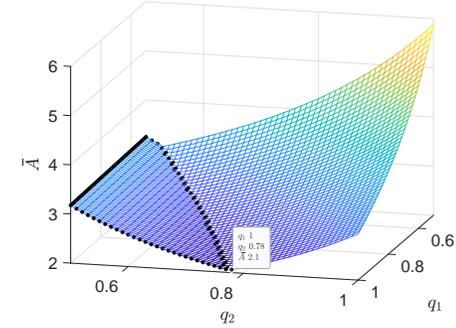}}
\caption{Average AoA for the infinite-sized battery for the second setup. The minimum $\bar{A}^*=2.1$ is achieved by $q^*_1=1$ and $q^*_2=0.78$.}
\label{Second_Set_AoA}
\end{figure}

Figs. \ref{First_set_AoA} and \ref{Second_Set_AoA} illustrate the average AoA for the case of an infinite-sized battery versus $q_1$ and $q_2$, for the first and the second setup, respectively. 
\vspace{0pt}
The minimum points and their values validate the analytical findings in (\ref{Opt_point_Ave_AoA}) and (\ref{Opt_value_Ave_AoA}). In Fig. \ref{First_set_AoA}, related to the first setup, we see that the minimum point is when $q^*_1=q^*_2=1$, which means that it is optimal to activate both transmitters.

In Fig. \ref{Second_Set_AoA}, related to the second setup in which interference can degrade significantly the success probability for the data transmission, we have $q^*_1=1$ and $q^*_2=0.78$, this is expected since we need to allow for a period of silence for the device transmitting power to not degrade the transmission of the data node. We cannot lower too much $q_2$ since then we may receive data without sufficient energy to perform the actions.
\begin{figure}[h]
\centerline{\includegraphics[scale=0.42]{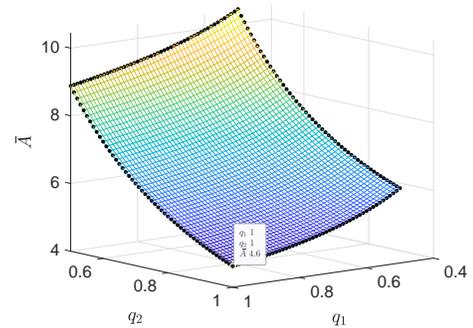}}
\vspace{-10pt}
\caption{Average AoA for the finite-sized battery with size of $m=1$, related to the first setup. The minimum $\bar{A}^*=4.6$ is achieved by $q^*_1=q^*_2=1$.}
\label{Finite1_First_set_AoA}
\end{figure}\

\begin{figure}[h]
\centerline{\includegraphics[scale=0.42]{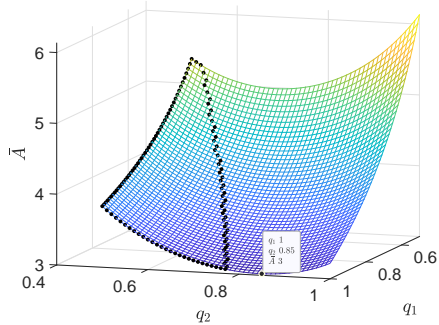}}
\vspace{-10pt}
\caption{Average AoA for the battery with size of $m=1$, related to the second setup. The minimum $\bar{A}^*=3$ is achieved by $q^*_1=1$ and $q^*_2=0.85$.}
\label{Finite1_Second_Set_AoA}
\end{figure}

In Figs. \ref{Finite1_First_set_AoA} and \ref{Finite1_Second_Set_AoA} we consider the case of having a battery of size $m=1$, for the first and the second setup, respectively. 
Note that the minimum values for the finite-sized battery case are obtained by exhaustive search. Comparing Figs. \ref{Finite1_First_set_AoA} and \ref{First_set_AoA}, related to the first setup, there is no difference in the optimum point $[q^*_1, q^*_2]$. From a design perspective, a small battery is sufficient for the optimal operation of that setup.

However, comparing Figs. \ref{Finite1_Second_Set_AoA} and \ref{Second_Set_AoA}, related to the second setup, we observe that as the battery size becomes smaller, the provision of energy is more critical. This is because the energy packets have a higher probability of successful transmission compared to data when both transmitters are active, and at the same time, the battery cannot store them due to low capacity. The battery may frequently be empty thus, many actions have to occur by consuming energy packets that have just arrived. This has also an impact on the minimum average AoA. 

As the battery size becomes larger, $q^*_2$ becomes smaller, since the energy packets can be stored and utilized later. Asymptotically, the optimal point tends to the border of the energy-limited area.

\section{Conclusion}
In this work, we defined a new metric termed Age of Actuation (AoA) which is more general than AoI and relevant to goal-oriented semantics-aware communications. Furthermore, it provides a link to control systems and multiple applications. AoA was studied in a system where a transmitter monitors a process and transmits status updates to a receiver. The receiver is a battery-powered device charged by a secondary transmitter dedicated to wireless energy transmission, and it performs actions based on the received updates. Furthermore, we studied two optimization problems that reveal the differences between AoA and AoI. This work opens up paths towards the study of more general and important problems, such as the incorporation of the proposed metric in practical systems, the characterization of its distribution, and the consideration of dynamic optimization.

\bibliographystyle{ieeetr}
\bibliography{bibliography.bib}

\begin{thebibliography}{10}

\bibitem{kountouris2021semantics}
M.~Kountouris and N.~Pappas, ``Semantics-empowered communication for networked
  intelligent systems,'' {\em IEEE Comm. Mag.}, vol.~59, no.~6, 2021.

\bibitem{gunduz2022beyond}
D.~Gunduz, Z.~Qin, I.~E. Aguerri, H.~S. Dhillon, Z.~Yang, A.~Yener, K.~K. Wong,
  and C.-B. Chae, ``Beyond transmitting bits: Context, semantics, and
  task-oriented communications,'' {\em arXiv:2207.09353}, 2022.

\bibitem{Timing6G}
P.~Popovski, F.~Chiariotti, K.~Huang, A.~E. Kalør, M.~Kountouris, N.~Pappas,
  and B.~Soret, ``A perspective on time toward wireless 6g,'' {\em Proceedings
  of the IEEE}, vol.~110, no.~8, pp.~1116--1146, 2022.

\bibitem{TutorialYates}
R.~D. Yates, Y.~Sun, D.~R. Brown, S.~K. Kaul, E.~Modiano, and S.~Ulukus, ``Age
  of information: An introduction and survey,'' {\em IEEE J. Sel. Areas
  Commun.}, vol.~39, no.~5, pp.~1183--1210, 2021.

\bibitem{Pappas2023age}
N.~Pappas, M.~Abd-Elmagid, B.~Zhou, W.~Saad, and H.~Dhillon, {\em Age of
  Information: Foundations and Applications}.
\newblock Cambridge University Press, Feb. 2023.

\bibitem{varshney2008transporting}
L.~R. Varshney, ``Transporting information and energy simultaneously,'' in {\em
  IEEE ISIT}, 2008.

\bibitem{YatesISIT2015}
R.~D. Yates, ``Lazy is timely: Status updates by an energy harvesting source,''
  in {\em IEEE ISIT}, 2015.

\bibitem{feng2018optimal}
S.~Feng and J.~Yang, ``Optimal status updating for an energy harvesting sensor
  with a noisy channel,'' in {\em IEEE INFOCOM WKSHPS}, 2018.

\bibitem{zheng2019closed}
X.~Zheng, S.~Zhou, Z.~Jiang, and Z.~Niu, ``Closed-form analysis of non-linear
  age of information in status updates with an energy harvesting transmitter,''
  {\em IEEE T. on Wireless Comm.}, vol.~18, no.~8, 2019.

\bibitem{bacingolu2019optimal}
B.~T. Bacinoglu, Y.~Sun, E.~Uysal, and V.~Mutlu, ``Optimal status updating with
  a finite-battery energy harvesting source,'' {\em Jour. of Comm. and
  Networks}, vol.~21, no.~3, 2019.

\bibitem{ArafaTWC2019}
A.~Arafa and S.~Ulukus, ``Timely updates in energy harvesting two-hop networks:
  Offline and online policies,'' {\em IEEE Transactions on Wireless
  Communications}, vol.~18, no.~8, pp.~4017--4030, 2019.

\bibitem{elmagid2022age}
M.~A. Abd-Elmagid and H.~S. Dhillon, ``Age of information in multi-source
  updating systems powered by energy harvesting,'' {\em IEEE Jour. on Selected
  Areas in Information Theory}, vol.~3, no.~1, 2022.

\bibitem{pappas2020average}
N.~Pappas, Z.~Chen, and M.~Hatami, ``Average aoi of cached status updates for a
  process monitored by an energy harvesting sensor,'' in {\em 54th CISS}, 2020.

\bibitem{hatami2021aoi}
M.~Hatami, M.~Leinonen, and M.~Codreanu, ``Aoi minimization in status update
  control with energy harvesting sensors,'' {\em IEEE T. on Comm.}, vol.~69,
  no.~12, 2021.

\bibitem{chen2021optimization}
Z.~Chen, N.~Pappas, E.~Björnson, and E.~G. Larsson, ``Optimizing information
  freshness in a multiple access channel with heterogeneous devices,'' {\em
  IEEE Open Jour. of the Comm. Society}, vol.~2, 2021.

\bibitem{CeranAoI2019}
E.~T. Ceran, D.~Gündüz, and A.~György, ``Reinforcement learning to minimize
  age of information with an energy harvesting sensor with harq and sensing
  cost,'' in {\em IEEE INFOCOM Workshops}, 2019.

\bibitem{xie2022age}
M.~Xie, Q.~Cao, M.~Zhou, and X.~Jia, ``Age of information for preemptive
  transmission in dual-sensor networks with energy harvesting,'' in {\em IEEE
  VTC-Spring)}, 2022.

\bibitem{krikidis2019average}
I.~Krikidis, ``Average age of information in wireless powered sensor
  networks,'' {\em IEEE Wireless Comm. Letters}, vol.~8, no.~2, 2019.

\bibitem{khorsandmanesh2021average}
Y.~Khorsandmanesh, M.~J. Emadi, and I.~Krikidis, ``Average peak age of
  information analysis for wireless powered cooperative networks,'' {\em IEEE
  T. on Cognitive Comm. and Networking}, vol.~7, no.~4, 2021.

\bibitem{abd2020aoi}
M.~A. Abd-Elmagid, H.~S. Dhillon, and N.~Pappas, ``Aoi-optimal joint sampling
  and updating for wireless powered communication systems,'' {\em IEEE T. on
  Vehicular Tech.}, vol.~69, no.~11, 2020.

\bibitem{leng2019ageof}
S.~Leng, X.~Ni, and A.~Yener, ``Age of information for wireless energy
  harvesting secondary users in cognitive radio networks,'' in {\em IEEE MASS},
  2019.

\bibitem{Abd-ElmagidTCOM2020}
M.~A. Abd-Elmagid, H.~S. Dhillon, and N.~Pappas, ``A reinforcement learning
  framework for optimizing age of information in rf-powered communication
  systems,'' {\em IEEE Transactions on Communications}, vol.~68, no.~8, 2020.

\bibitem{zhang2013mimo}
R.~Zhang and C.~K. Ho, ``Mimo broadcasting for simultaneous wireless
  information and power transfer,'' {\em IEEE T. on Wireless Comm.}, vol.~12,
  no.~5, 2013.

\bibitem{nguyen2008optimization}
G.~D. Nguyen, S.~Kompella, J.~E. Wieselthier, and A.~Ephremides, ``Optimization
  of transmission schedules in capture-based wireless networks,'' in {\em IEEE
  MILCOM}, 2008.

\bibitem{fountoulakis2022information}
E.~Fountoulakis, T.~Charalambous, N.~Nomikos, A.~Ephremides, and N.~Pappas,
  ``Information freshness and packet drop rate interplay in a two-user
  multi-access channel,'' {\em Journal of Communications and Networks},
  vol.~24, no.~3, 2022.

\bibitem{srikant2013communication}
R.~Srikant and L.~Ying, {\em Communication networks: an optimization, control,
  and stochastic networks perspective}.
\newblock Cambridge University Press, 2013.

\end{thebibliography}
\end{document}